# Lifetime Prediction of 1550 nm DFB Laser using Machine learning Techniques

Khouloud Abdelli[1,2], Danish Rafique[1], Helmut Grießer[1], and Stephan Pachnicke[2]

[1]*ADVA Optical Networking SE, Fraunhoferstr. 9a, 82152 Munich/Martinsried, Germany*
[2] *Christian-Albrechts-Universität zu Kiel, Kaiserstr. 2, 24143 Kiel, Germany*
*E-mail: KAbdelli@advaoptical.com*

**Abstract:** A novel approach based on an artificial neural network (ANN) for lifetime prediction of 1.55 µm InGaAsP MQW-DFB laser diodes is presented. It outperforms the conventional lifetime projection using accelerated aging tests. © 2020 The Author(s)

1. **Introduction**

Distributed Feedback (DFB) lasers are one of the main light sources of long-haul and high-bite rate optical fiber transmission systems requiring high performance and reliability. Hence, there have been several research undertakings to improve the characteristics of such lasers [1].
Laser reliability is strongly dependent on the device characteristics such as current versus optical power behavior and its temperature dependence as well as the physical device degradation [1]. Traditionally, laser reliability is characterized by monitoring either the increase in operating current $I_{op}$ or the decrease in optical power $P_{op}$. The laser lifetime is defined as the time at which the degradation parameters $I_{op}$ or $P_{op}$ reach their maximum acceptable threshold. Depending on the operating conditions, accelerated aging tests are performed, according to Telcordia GR-468 CORE requirements, for the reliability evaluation, allowing the reduction of the test time and long-term reliability prediction. These tests must be conducted on many devices aged during several thousands of hours at higher than normal temperature and operating currents. Note that the typical required laser mean-time-to-failure (MTTF) in optical communication system is $> 10^5$ hours [2]. Nonetheless, the aging tests [1] are costly, time consuming and imprecise due to the equipment measurement and control instability. Furthermore, the conventional method based on a single degradation parameter $I_{op}$ or $P_{op}$ could overestimate the actual laser lifetime. Multiple empirical degradation models based on several laser parameters are proposed in [2]. Recently, new approaches based on machine learning (ML) techniques achieving higher reliability and confidence levels have been proposed to improve the laser reliability via a generic neural network model. In this regard, we investigated a data-driven fault detection model based on Long Short-Term Memory (LSTM) to detect the different laser failure modes [3].
In this paper, we present an ANN-based model for predicting with higher accuracy the MTTF of a 1.55 µm InGaAsP MQW-DFB laser under different operating conditions. The proposed model takes the different laser characteristics as the inputs and MTTF as the output. Synthetic data is used for training and validation of the model. The overall performance evaluation of the model indicates a mean square error of 0.2 years. Our analysis shows that ANN outperforms the conventional laser lifetime prediction method both in terms of accuracy and the application to unseen operating conditions.

2. **Case Study**

This paper develops a lifetime model using ANN able to predict the MTTF of a laser during its operation given different monitored laser parameters. By modelling the dependency between the laser lifetime and the impact of the different laser characteristics on its reliability, the proposed model can accurately predict the MTTF even under unseen operating conditions. Thus, saving the time and the costs required to carry out the accelerated aging tests for the laser quantification under different conditions. Note that typically such an ML model would be trained using experimentally derived MTTF values, however due to lack of an extensive dataset, we considered an analytical approach for synthetic generation of the MTTF values. The results presented thus show the feasibility of our approach in terms of MTTF prediction using ANN for lasers with vastly different operating conditions.

*2.1. Data Generation*

In order to train the ML model, synthetic data was generated for low power ($P_{op} \leq 10\ mW$) InGaAsP MQW-DFB lasers operating at a $\lambda$ range from 1.53 to 1.57 µm in the case temperature $T_c$ range of -40 °C to 85 °C with side mode suppression ratio SMSR of more than 35 dB. A real laser datasheet [4] is used to model the different laser electro-optical characteristics namely current threshold $I_{th}$, slope efficiency SE, voltage V and $\lambda$. The performance curves of the modeled parameters are shown in Fig. 1.

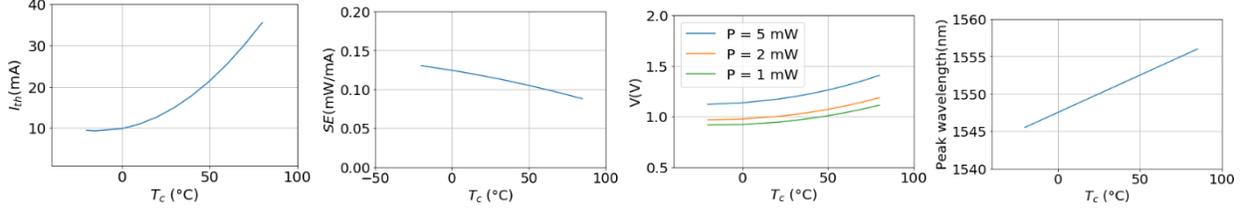

*Figure 1 a). $I_{th}$ versus $T_c$     b) SE versus $T_c$     c) V versus $T_c$ , $P_{op}$     d) $\lambda$ versus $T_c$*

For given values of $T_c$ and $P_{op}$, randomly selected from uniform distributions, the different laser parameters namely $I_{th}$, SE, V, $\lambda$, $\eta$ [5] and junction temperature $T_j$ [5] are calculated. The $MTTF_2$ of the obtained parameters is estimated using a model for two stress terms temperature and the forward current [5] expressed as

$$MTTF_2 = MTTF_1 \left(\frac{I_1}{I_2}\right)^2 \exp\left(\frac{E_a}{k_B}\left(\frac{1}{T_{j,2}} - \frac{1}{T_{j,1}}\right)\right) \quad (1)$$

Where $E_a$ is the activation energy for the device in units of eV, $k_B$ is Boltzmann's constant, $T_{j,1}$ and $T_{j,2}$ are different junction temperatures in units of Kelvin, $I_1$ and $I_2$ are the corresponding operating currents at $T_{j,1}$ and $T_{j,2}$ respectively. In our model, we assumed the aforementioned parameters to be fixed, and used the experimentally identified $MTTF_1$, denoted by MTTF, ($7 \times 10^5$ hours), based on accelerated aging test under stressed conditions ($P_{op} = 10\ mW, T_c = 50\ °C$).

The process of the data generation is shown in Fig. 2.

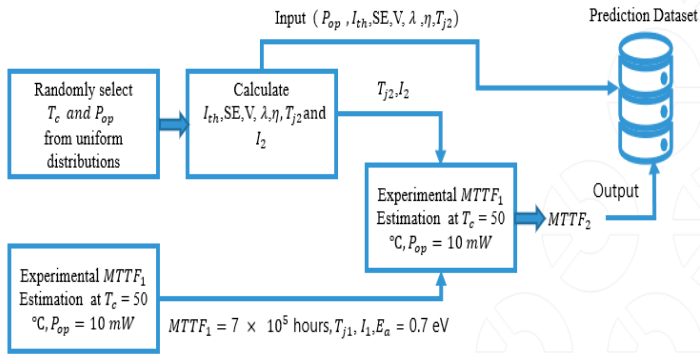 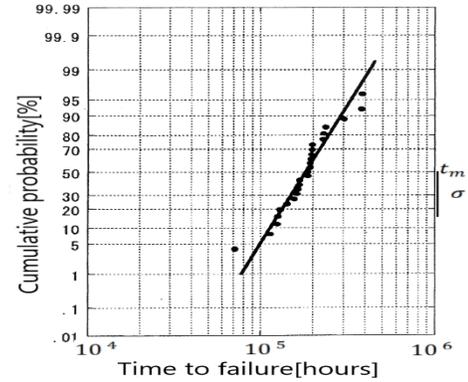

*Figure 2. Dataset Generation Process*      *Figure 3. Cumulative failure versus failure times*

### 2.2. Artificial Neural Network Model

Fed with normalized values of the inputs $P_{op}$, $I_{th}$, $\eta$, SE, V, $\lambda$ and $T_j$, the ANN model was trained using the back-propagation algorithm. A mean square error cross-entropy function was used as the loss function to update the weights of the model based on the error between the predicted and the desired output. Furthermore, the hyperparameter tuning of the model led to the selection of the Adam optimizer and the number of hidden layers of 2.

The accuracy of the model prediction was quantified by using two regression evaluation metrics, i.e mean square error (mse) and the value of scoring function $s$ [6] which can be expressed as function of $h_i$, denoting the difference between the ith predicted value and ith true value as

$$s = \begin{cases} \sum_{i=1}^{N}(e^{\frac{-h_i}{13}} - 1) & \text{for } h_i < 0 \\ \\ \sum_{i=1}^{N}(e^{\frac{h_i}{10}} - 1). & \text{for } h_i \geqslant 0 \end{cases} \quad (2)$$

The scoring function measures the error of estimation by penalizing overestimated values more than the underestimated values as the overestimated values lead to device failure resulting in higher costs.

### 2.3. Conventional Lifetime Prediction Method

Accelerated aging tests for twenty-five 1.55 µm lasers (also DFB laser as assumed in the model) conducted for 5000 hours under constant output optical power of 10 mW at 70 °C were carried out. The cumulative percent failures versus the failure times of the devices estimated by linear extrapolation of change of $I_{op}$ up to 50% are plotted on a lognormal probability scale depicted in figure 3 which yielded $1.8 \times 10^5$ hours for median lifetime $t_m$, defined as the point of the fitted line where 50% of devices failed the test, and 0.4 for the standard deviation $\sigma$ calculated as $\ln(t_m/t_1)$, where $t_1$ is the time corresponding to a cumulative failure of 16% [7]. Consequently, the $MTTF_1$ is calculated as [7]

$$MTTF_1 = t_m \exp\left(\frac{\sigma^2}{2}\right). \tag{3}$$

Given a $MTTF_1$ of $1.9 \times 10^5$ hours, the $MTTF_2$ of the same device at a different junction temperature $T_{j,2}$ can be estimated using the Arrhenius model[7].

## 3. Results and Discussions

To evaluate the accuracy and the generalization capabilities of the developed model, the ANN model was tested to predict the MTTF on an unseen test dataset. The summary of the ANN model performance is shown in figure 4. The ANN model achieved smaller MSE of 0.2 years and smaller score function of 17. The scoring function plot, illustrating the individual scoring function of each test dataset point as function of $h_i$ denoting the prediction error, as depicted in figure 4, shows that the ANN model could overestimate the MTTF up to 1.2 years and could underestimate it up to 1.3 years. The ANN model was compared in terms of MSE and scoring function with two standard regression ML algorithms namely Random Forest (RF) and Gradient Boosting Machine (GBM) trained with the same dataset used to feed the ANN model. The results demonstrated that ANN significantly outperformed the other ML algorithms by achieving smallest MSE as well as smallest score function value

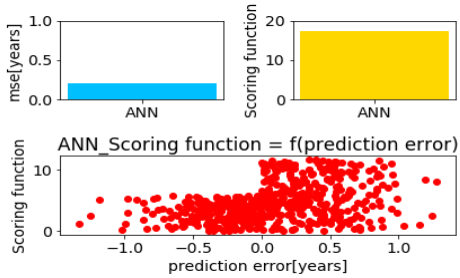 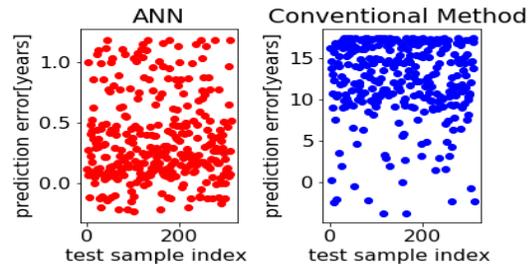

*Figure 4. ANN performance evaluation summary.*     *Figure 5. Comparison of ANN model with conventional method*

In order to compare the performance of the ANN model with the conventional laser lifetime prediction method, a test dataset, containing different laser characteristics estimated under similar operating conditions like the accelerated aging tests (same optical power $P_{op} = 10\ mW$, different temperatures), was generated and used. The results of the comparison, as shown in figure 5, highlight that the ANN model outperformed the conventional approach in terms of prediction error. The ANN model was able to predict correctly the MTTF with prediction error less than 1.12 years. Whereas the traditional method predicted the MTTF with error in the range of 3.8 to 17 years.

## 4. Conclusions

This paper proposes a ML approach based on ANN for laser MTTF prediction. Synthetic data including different electro-optical laser characteristics is used to evaluate the prediction accuracy of the proposed model. The results show that ANN outperforms the other ML algorithms namely RF and GBM in terms of MSE and scoring function as well as the conventional method based on accelerated aging tests.

Future work will focus on the accuracy improvement of the proposed model by including other laser characteristics influencing its reliability such as SMSR, Relative Intensity Noise, and resistance during the training of the model.

## 5. References

[1] T. Ikegami., M. Fukuda., M. Suzuki., "Considerations on the Degradation of DFB Lasers", in Christou A., Unger B.A. (eds) Semiconductor Device Reliability. NATO ASI Series (Series E: Applied Sciences), vol 175. Springer, Dordrech, 1990
[2] Nam Hwang, Seung-Goo Kang, Hee-Tae Lee, Min-Kyu Song, Kwang-Eui Pyun, "Degradation Models and Lifetime Projections of InGaAs/InP MQW-DFB Laser Diodes for High Speed Optica," Proceedings of International Reliability Physics Symposium, Dallas, TX, USA, 1996, pp. 195.
[3] K. Abdelli, D. Rafique, S. Pachnicke "Machine Learning Based Laser Failure Mode Detection", ICTON, 2019
[4] 1500 nm InGaAsP MQW-DFB laser datasheet available at http://www.cel.com/pdf/datasheets/ndl7705.pdf
[5] Li, Yajun, "Lifetime estimation of InGaAlP lasers under different operating conditions", SPIE Proc. Vol. 5878, 2005.
[6] A. Saxena, K. Goebel, "PHM08 Challenge data set," NASA Ames Prognostics Data Repository (http://ti.arc.nasa.gov/project/prognostic-data-repository), NASA Ames Research Center, Moffett Field, CA, 2008.
[7] Han, Jae-Ho ,Park, Sung-Woong. (2007),"Reliability of loss-coupled 1.55 μm DFB laser diode with automatically buried absorptive InAsP layer", Microwave and Optical Technology Letters, 2007.